\documentclass[sigconf]{acmart}
\AtBeginDocument{%
  }

\setcopyright{acmlicensed}
\copyrightyear{2018}
\acmYear{2018}
\acmDOI{XXXXXXX.XXXXXXX}
\acmConference[FSE '26]
{The ACM International Conference on the Foundations of Software Engineering (FSE 2026)}
{July 5--9, 2026}
{Montréal, Canada}

\acmISBN{978-1-4503-XXXX-X/2018/06}

\usepackage{hyperref} 
\usepackage{xspace}
\usepackage{color}
\usepackage{xcolor}
\usepackage{fancybox}
\usepackage{fancyhdr}
\usepackage{ifthen} 
\usepackage{paralist} 

\usepackage[T1]{fontenc}
\usepackage{graphicx} 
\usepackage{courier}
\usepackage{listings} 
\usepackage{multirow}

\usepackage[caption=false]{subfig} 
\usepackage{makecell} 

\definecolor{keywordcolor}{rgb}{0.0, 0.2, 0.6} 
\definecolor{commentcolor}{rgb}{0.3, 0.6, 0.3} 
\definecolor{stringcolor}{rgb}{0.2, 0.5, 0.2} 
\definecolor{classcolor}{rgb}{0.7, 0.5, 0.0} 
\definecolor{annotationcolor}{rgb}{0.6, 0.3, 0.0} 
\definecolor{numbercolor}{rgb}{0.4, 0.4, 0.4} 

\lstdefinestyle{javacode}{
  language=Java,
  basicstyle=\ttfamily\footnotesize\color{black}, 
  keywordstyle=\color{keywordcolor}, 
  commentstyle=\color{brown}\itshape, 
  stringstyle=\color{stringcolor}, 
  classoffset=1, 
  morekeywords={UploadImage}, 
  keywordstyle=[1]\color{classcolor}, 
  keywordstyle=[2]\color{annotationcolor}, 
  morekeywords=[2]{@Test}, 
  numbers=left,
  numberstyle=\tiny\color{numbercolor}, 
  stepnumber=1,
  frame=single,
  tabsize=4,
  showstringspaces=false,
  breaklines=true,
  captionpos=b,
  numbersep=10pt, 
  framexleftmargin=15pt, 
  xleftmargin=30pt, 
}

\lstdefinelanguage{json}{
    basicstyle=\ttfamily\footnotesize\color{black}, 
    numbers=left, 
    numberstyle=\tiny\color{numbercolor}, 
    stepnumber=1, 
    numbersep=10pt, 
    showstringspaces=false, 
    breaklines=true, 
    frame=single, 
    tabsize=4, 
    framexleftmargin=15pt, 
    xleftmargin=30pt, 
    captionpos=b, 
    literate=
     *{0}{{{\color{blue}0}}}{1}%
      {1}{{{\color{blue}1}}}{1}%
      {2}{{{\color{blue}2}}}{1}%
      {3}{{{\color{blue}3}}}{1}%
      {4}{{{\color{blue}4}}}{1}%
      {5}{{{\color{blue}5}}}{1}%
      {6}{{{\color{blue}6}}}{1}%
      {7}{{{\color{blue}7}}}{1}%
      {8}{{{\color{blue}8}}}{1}%
      {9}{{{\color{blue}9}}}{1}%
      {:}{{{\bfseries\color{black}:}}}{1}%
      {,}{{{\bfseries\color{black},}}}{1}%
      {\{}{{{\color{black}\{}}}{1}%
      {\}}{{{\color{black}\}}}}{1}%
      {[}{{{\color{black}[}}}{1}%
      {]}{{{\color{black}]}}}{1}%
}

\lstdefinestyle{csharpcode}{
  language=[Sharp]C,
  basicstyle=\ttfamily\footnotesize\color{black},
  keywordstyle=\color{keywordcolor},
  commentstyle=\color{brown}\itshape,
  stringstyle=\color{stringcolor},
  numbers=left,
  numberstyle=\tiny\color{numbercolor},
  stepnumber=1,
  frame=single,
  tabsize=4,
  showstringspaces=false,
  breaklines=true,
  captionpos=b,
  numbersep=10pt,
  framexleftmargin=15pt,
  xleftmargin=30pt,
  morekeywords={async,await,var,dynamic,get,set,partial,record,init,yield,from,select,where,orderby,group,into,join,let,equals},
  classoffset=1,
  morekeywords=[1]{Test,Fact,Theory,SetUp,TearDown,OneTimeSetUp,OneTimeTearDown,Category},
  keywordstyle=[1]\color{annotationcolor},
  classoffset=0
}

\lstdefinelanguage{gherkin}{
  sensitive=true,
  morecomment=[l]{\#},
  morestring=[b]",
  morekeywords={
    Feature,Background,Scenario,ScenarioOutline,Examples,Rule,
    Given,When,Then,And,But
  }
}

\lstdefinestyle{gherkincode}{
  language=gherkin,
  basicstyle=\ttfamily\footnotesize\color{black},
  keywordstyle=\color{keywordcolor}\bfseries,
  commentstyle=\color{brown}\itshape,
  stringstyle=\color{stringcolor},
  numbers=left,
  numberstyle=\tiny\color{numbercolor},
  stepnumber=1,
  frame=single,
  tabsize=2,
  showstringspaces=false,
  breaklines=true,
  captionpos=b,
  numbersep=10pt,
  framexleftmargin=15pt,
  xleftmargin=30pt
}

\begin{document}

\title{Prompt-Based REST API Test Amplification in Industry: An Experience Report}

\author{Tolgahan Bardakci}
\authornote{Corresponding Author}
\email{tolgahan.bardakci@uantwerpen.be}
\orcid{0009-0007-1136-2065}
\affiliation{%
  \institution{University of Antwerp and Flanders Make}
  \city{Antwerp}
  \state{Antwerp}
  \country{Belgium}
}

\author{Andreas Faes}
\email{andreas.faes@katoennatie.com}
\orcid{0009-0002-4402-5552}
\affiliation{%
  \institution{Katoen Natie}
  \city{Antwerp}
  \state{Antwerp}
  \country{Belgium}
}

\author{Mutlu Beyaz{\i}t}
\email{mutlu.beyazit@uantwerpen.be}
\orcid{0000-0003-2714-8155}
\affiliation{%
  \institution{University of Antwerp and Flanders Make}
  \city{Antwerp}
  \state{Antwerp}
  \country{Belgium}
}

\author{Serge Demeyer}
\email{serge.demeyer@uantwerpen.be}
\orcid{0000-0002-4463-2945}
\affiliation{%
  \institution{University of Antwerp and Flanders Make}
  \city{Antwerp}
  \state{Antwerp}
  \country{Belgium}
}

\renewcommand{\shortauthors}{Bardakci et al.}

\begin{abstract}
Large Language Models (LLMs) are increasingly used to support software testing tasks, yet there is little evidence of their effectiveness for REST API testing in industrial settings.
To address this gap, we replicate our earlier work on LLM-based REST API test amplification within an industrial context at one of the largest logistics companies in Belgium.
We apply LLM-based test amplification to six representative endpoints of a production microservice embedded in a large-scale, security-sensitive system, where there is in-depth complexity in authentication, stateful behavior, and organizational constraints.
Our experience shows that LLM-based test amplification remains practically useful in industry by increasing coverage and revealing various observations and anomalies.
\end{abstract}

\ccsdesc[500]{Software and its engineering}
\ccsdesc[500]{Software and its engineering~Functionality}
\ccsdesc[500]{Software and its engineering~Software testing and debugging}
\ccsdesc[500]{Computing methodologies~Artificial intelligence}

\keywords{Software Testing, Test Amplification, Large Language Models}




\maketitle

\section{Introduction}
REST APIs (Representational State Transfer Application Programming Interfaces) have become a fundamental building block of modern software systems, enabling communication between distributed services in cloud-native and microservice-based architectures.
In industrial settings, REST APIs often serve as the primary integration layer between independently developed components, external partners, and customer-facing applications.
As a result, failures at the API level can have severe operational and business consequences, ranging from service outages to data inconsistencies and security incidents.
Therefore, ensuring the correctness and robustness of REST APIs is a critical concern in industrial software engineering practice and places strong demands on effective, maintainable API-level testing.
However, this comes with challenges due to the number of input combinations, diverse data dependencies, and independently evolving service endpoints.
Thus, testers search for a small fraction of interactions that will expose bugs, resulting in the "needle in the haystack" phenomenon.

Test amplification is a likely candidate to solve this problem.
It is an umbrella term for the various activities that analyze and operate on existing test suites, such as augmentation, optimization, enrichment, or refactoring, and it has been shown to be effective ~\cite{test_amplification_definition}.
Rather than generating tests from scratch, test amplification operates on existing test cases and augments them with additional inputs, assertions, or execution paths, thereby preserving the original test intent and structure.
During this process, they retain contextual relevance by reusing and refining existing tests.
Thus, outputs align more closely with the way developers write and maintain the tests.
This aspect of test amplification makes it especially attractive in industrial settings.

Recent advances in Large Language Models (LLMs) have sparked growing interest in their use for supporting software engineering tasks, including program understanding, code generation, and testing~\cite{zhang2024surveylargelanguagemodels, hou2024LLMSforSE}.
In the context of software testing, LLMs are increasingly being explored to reduce manual effort by generating, extending, or improving test cases based on existing code and specifications~\cite{unit_test_amplification-meta, nooyens2025agenticamplification}.
Prior studies have shown that LLMs can generate readable, syntactically correct test code and assist developers in strengthening test suites with limited human input~\cite{biagiola2025ReadabilityLLMs, besjes2025agenticllmsrestapi}.
Their use in testing REST APIs has also been explored, suggesting that LLMs may offer practical opportunities to improve test effectiveness, including at the API level.

Despite the growing work on applying LLMs to software testing, existing evaluations are largely based on open-source projects, synthetic benchmarks, or controlled experimental settings~\cite{augusto2025largelanguagemodelssoftware}.
While such studies are valuable for establishing feasibility, they often abstract away key characteristics of industrial systems, such as authentication mechanisms, stateful interactions, organizational conventions, and operational constraints~\cite{nooyens2025agenticamplification}.
As a result, it remains unclear to what extent insights and benefits observed in open source example analyses carry over to real-world REST API testing.
This lack of industrial evidence limits both practitioners’ confidence in adopting LLM-based testing approaches and researchers’ understanding of their practical limitations.

In earlier work, we investigated the use of out-of-the-box LLMs for amplifying REST API test suites in a controlled, open-source setting~\cite{bardakci2025testamplificationrestapis}.
Starting from an existing happy-path API test, we showed that LLMs can generate additional test cases that improve structural API coverage and, in some cases, expose anomalies.
Our evaluation, conducted on an open-source REST API, focused on coverage, readability, and post-processing effort.
It demonstrated that prompt design plays a significant role in the quality of the amplified tests.
While these results provided an initial indication of the potential of LLM-based test amplification, they were intentionally limited to a non-industrial context, leaving open questions about applicability under real-world constraints.

In this paper, we address the lack of industrial evidence by studying LLM-based test amplification in a real-world REST API testing context.
We revisit the approach explored in our earlier work~\cite{bardakci2025testamplificationrestapis} and apply it to a production microservice within a large-scale industrial system.
The system belongs to one of the largest logistics companies in Belgium.
We focus on a small number of carefully selected endpoints that exhibit realistic industrial complexity.
Our study aims to assess both the practical usefulness and the limitations of LLM-based test amplification beyond standard settings.

Consequently, this paper contributes to the body of knowledge 
\begin{itemize}
    \item by conducting an industrial replication study of LLM-based REST API test amplification, and
    \item deriving a set of lessons learned and practical recommendations for practitioners considering LLM-based test amplification in real-world API testing environments.
\end{itemize}

To guide our investigation, we address the following research questions:
\newcommand*{\RQCoverageImpact} {\textit{To what extent does LLM-based test amplification improve structural API coverage in an industrial REST API?}}
\newcommand*{\RQEffortAndPracticalCost} {\textit{What level of human effort is required to integrate LLM-\allowbreak\\amplified tests into an existing industrial test suite?}}
\newcommand*{\RQObservationsAndLimitations} {\textit{What practical challenges and anomalies arise when applying LLM-based test amplification in an industrial setting?}}
\begin{itemize}
    \item RQ1: \RQCoverageImpact
    \item RQ2: \RQEffortAndPracticalCost
    \item RQ3: \RQObservationsAndLimitations
\end{itemize}

The remainder of this paper is structured as follows.
Section~\ref{section:IndustrialContext} and Section ~\ref{EvaluationSetup} describe the industrial system under study and the experimental design, followed by the presentation of results in Section~\ref{sec:Results}.
We then discuss lessons learned and threats to validity before concluding the paper.

\section{Related Work}
\paragraph{LLM-based Test Generation and Amplification.}
Recent years have seen growing interest in applying LLMs to software testing tasks, particularly for automated test generation and test improvement.
Prior work shows that LLMs can generate syntactically correct, readable, and aligned unit tests when provided with sufficient context \cite{schäfer2024empiricalevaluationusinglarge, yang2024evaluationlargelanguagemodels}.
Within this space, test amplification has emerged as a complementary approach that focuses on strengthening existing tests rather than generating new ones from scratch, thereby preserving test intent and architectural consistency \cite{test_amplification_definition, test_amplification_ampyfier}.
While several studies demonstrate the feasibility of LLM-based test amplification in a controlled environment for unit tests, there are few studies that show that test amplification can improve test effectiveness by increasing coverage and exposing corner cases in REST API testing, while remaining maintainable~\cite{bardakci2025testamplificationrestapis}~\cite{nooyens2025agenticamplification}.
These studies show that LLMs can generate readable, executable amplified tests, improve coverage, and occasionally reveal anomalies.
They also show that the prompt design is very important.
Unlike existing work, our work aims to scrutinize the applicability of REST API test amplification under industrial constraints.

\paragraph{Industrial LLM Studies in Testing.}
A smaller but growing body of work reports on the use of LLMs for testing tasks in industrial environments.
These studies emphasize integration into existing development workflows, human-in-the-loop validation, and organizational constraints \cite{unit_test_amplification-meta}. 
Findings consistently show that while LLMs can improve developer productivity and code coverage, their outputs require human review and adaptation to align with project-specific conventions.
Compared to these studies, our work extends industrial evidence to REST API testing and test amplification, and highlights challenges that arise specifically in integration-level testing scenarios.

\paragraph{REST API Testing and API Coverage.}
REST API testing has been extensively studied, with approaches ranging from functional testing frameworks to fuzzing tools \cite{golmohammadi2022testingrestfulapissurvey, atlidakis2019restler}.
Structural API coverage metrics derived from OpenAPI specifications have been proposed to quantify how thoroughly tests exercise an API without requiring access to internal implementations \cite{test_coverage_criteria}.
Existing tools and techniques primarily focus on generating test cases from API specifications or observed traffic, whereas fewer works address the amplification of existing API test suites.
Our work lies at the intersection of these research directions by evaluating LLM-based test amplification using API coverage metrics and validating the approach in an industrial REST API setting.

\section{Industrial Context and System Under Test}\label{section:IndustrialContext}
This section outlines the context of our industrial study, and also gives information about the system under test and the endpoints selected for the study.

\subsection{Company and Constraints}
The study was conducted in collaboration with a large industrial partner operating in the logistics domain.
The company is one of the largest logistics organizations in Belgium and operates complex, software-intensive systems that support core business processes such as shipment tracking, order management, routing, billing.
These systems are developed and maintained by multiple teams and are deployed across [cloud/on-premise/hybrid] environments, serving both internal stakeholders and external partners.

As is common in industrial settings, the study was subject to strict security, privacy, and compliance constraints.
All experiments were conducted in the development environment using controlled data.
In addition, organizational policies restricted the use of external tools and the disclosure of proprietary interfaces, requiring careful handling of API specifications, test artifacts, and experimental results.

Due to the size and complexity of the overall system, evaluating LLM-based test amplification across all services and endpoints was neither feasible nor desirable.
The company’s software ecosystem consists of hundreds of microservices, many of which interact through asynchronous workflows and shared infrastructure.
Testing the entire system would have introduced substantial confounding factors unrelated to test amplification itself, such as cross-team dependencies, environment instability, and deployment coordination.
Consequently, the study focuses on a single microservice and a limited set of representative endpoints, enabling a controlled yet realistic assessment of LLM-based test amplification under industrial constraints.

\subsection{Service Under Test}\label{ref:ServiceUnderTest}
The study focuses on a single production microservice that plays a central role in the company’s core business process, focusing on warehouse management.
The service is responsible for creating, updating, and querying domain entities and is invoked by multiple upstream systems, including internal services, external partners, and customer-facing applications.
As a result, failures or regressions in this service can directly affect business operations and service reliability.

From a testing perspective, the selected microservice is representative of typical industrial REST APIs.
It exposes a set of endpoints that require authentication and authorization, enforce business-specific validation rules, and exhibit state-dependent behavior across multiple requests.
In addition, the service adheres to strict organizational conventions regarding request schemas, error handling, and response formats, which are common in large-scale industrial systems but often absent from open-source example applications.

This microservice was chosen for the study because it satisfies three key criteria. First, it is business-critical and actively maintained, ensuring that test quality is a practical concern rather than an academic exercise.
Second, it exposes sufficient behavioral complexity to challenge automated test amplification techniques without relying on artificial or synthetic constructs.
Third, its scope allows controlled experimentation within the constraints of industrial development processes, enabling the application of LLM-based test amplification without disrupting ongoing development or deployment activities.
Additionally ---as for many private company systems--- LLM has no prior training of used cloud system or its artifacts.

\subsection{Endpoint Selection}
To ground the study in realistic usage scenarios, we selected six representative endpoints from the microservice under test.
The endpoints were chosen in collaboration with industrial developers based on the business need.
Typically, our industrial partner requires at least 80\% coverage for their services, including the service selected in this study.
Table~\ref{tab:endpoints} summarizes the selected endpoints and highlights their key traits relevant to test amplification.

\begin{table}[htbp]
\centering
\caption{Selected Endpoints and Their Characteristics\\
\footnotesize{\textit{Endpoint paths are anonymized due to confidentiality constraints.}}}
\label{tab:endpoints}
\begin{tabular}{l l l l l}
\hline
\textbf{ID} & \textbf{Endpoint Path} & \textbf{Method}  & \textbf{Payload}  \\
\hline
E1 & /... & GET     & Medium \\
E2 & /... & POST    & High   \\
E3 & /... & GET     & Low    \\
E4 & /... & PUT     & High   \\
E5 & /... & PATCH   & Medium \\
E6 & /... & PATCH   & Medium \\
\hline
\end{tabular}
\end{table}

All endpoints in the service employ authentication mechanisms and exhibit state-dependent behavior.
We estimate the request complexity of each endpoint based on the amount and structure of required request data, including authentication headers and request bodies.
Endpoints with higher request complexity are associated with a greater payload burden.

\section{Evaluation Setup}\label{EvaluationSetup}
Now that we introduced the industrial setting, we explain other elements that are important for our evaluation. 

\subsection{Method}

\emph{In our earlier work}~\cite{bardakci2025testamplificationrestapis}, we explored the use of out-of-the-box Large Language Models to amplify REST API tests in an open-source cloud application setting.
Starting from a manually written happy-path test, we prompted LLMs to generate additional test cases targeting boundary conditions and alternative execution paths.
The study was conducted on the PetStore application~\cite{SwaggerPetstore}, a widely used open-source cloud application.
The evaluation was carried out using structural API coverage metrics derived from Lopez et al.~\cite{test_coverage_criteria} and quality metrics such as readability and post-processing effort.
The results showed that LLMs can generate readable and executable amplified tests, improve coverage, and occasionally reveal anomalies.
Additionally, the outcomes show the importance of prompt design.

To enable a meaningful comparison with our prior study, we intentionally replicate the same experimental method without modification.
In particular, we reuse the same test amplification strategy, prompt structure, and evaluation objectives, and apply them directly to the industrial system under test. 
This design choice allows us to isolate the impact of the system context—open-source service versus industrial production service—while keeping all other experimental factors constant.

As in our previous work~\cite{bardakci2025testamplificationrestapis}, the amplification process starts from an existing manually written happy-path REST API test.
In the industrial study, the test scenario itself was authored by the research team, while the surrounding test framework and supporting components were derived from the company’s existing testing infrastructure.
This test serves as the seed input to the LLM, which is prompted to generate additional test cases to increase coverage, explore alternative execution paths, test boundary conditions, and handle erroneous inputs.
The amplified tests are then integrated into the existing test suite and executed against the system under test with minimal manual modification prior to execution.

We also reuse the same prompt design strategies explored in the prior study, including (i) providing the initial test code alone, (ii) augmenting the prompt with API documentation, and (iii) requesting a maximal number of amplified tests.
The goal of the amplification remains unchanged: to strengthen the existing test suite by increasing coverage, exercising different API behaviors, rather than to replace manually written tests or generate tests from scratch.


\subsection{Adaptations for Industry}
While the core method of test amplification was kept unchanged, several adaptations were necessary to apply the approach in an industrial setting.
These adaptations were driven by differences in technology stacks, development practices, and organizational constraints, rather than by changes to the amplification strategy itself.

A first adaptation concerns the test framework and programming language. 
In our prior study~\cite{bardakci2025testamplificationrestapis}, REST API tests were implemented in Java programming language using the REST Assured framework.
In contrast, the industrial system under test employs a Behavior-Driven Development (BDD) approach based on Gherkin specifications and C\# step definitions.
Consequently, LLM-based test amplification in this setting required generating both Gherkin feature files and corresponding C\# code.
This introduced an additional layer of complexity, as amplified tests needed to remain consistent across specification-level scenarios and executable step implementations.

In practice, not all generated C\# code was directly usable.
Parts of the generated step definitions duplicated functionality already available in the existing test code base, such as shared setup logic and reusable helper methods.
In these cases, we discarded the redundant generated code and manually linked the amplified Gherkin steps to existing step implementations.
This decision reflects common industrial testing practices, where reuse and consistency with established test infrastructure take precedence over fully generated code.

Finally, the choice of tooling (LLMs usage) is influenced by privacy and security constraints.
Whereas our previous study relied on cloud-based LLMs, the industrial replication is conducted using OpenAI's "ChatGPT 4o"~\cite{openai} integrated into the Visual Studio IDE~\cite{visualstudio}.
This setup complied with company policies regarding data exposure and ensured that source code and API specifications were not transmitted outside approved development environments.
Although this change affected the interaction modality with the LLM, it did not alter the underlying prompts or amplification goals.

\subsection{Metrics and Evaluation Criteria}

As in the previous study, we measure the impact of test amplification using structural API coverage metrics derived from the API specification~\cite{test_coverage_criteria}.
These metrics capture the extent to which the amplified test suite exercises different aspects of the REST API, such as paths, operations, parameters, request and response types, without requiring access to the service's internal implementation.
In addition to coverage, we assess the amount of post-processing effort required to integrate the generated tests into the existing test suite before the amplified tests can be adopted in practice.

\paragraph{Tooling for Coverage Measurement}

In our previous study, structural API coverage was computed using the open-source tool Restats~\cite{restats}.
When applying the same tool in the industrial context, we encountered several practical limitations that prevented its direct use, including parsing issues, incompatibilities with the test environment, stability issues, and integration challenges within the company’s CI pipeline.
These limitations highlight a common gap between research tooling and industrial environments.

To overcome these issues, we developed a lightweight coverage measurement tool, RESTCov~\cite{RESTCov_2025}, tailored to the industrial system under test, while the company independently developed an internal tool with similar functionality.
The difference is that the tool implemented by the company supports only a subset of the REST API coverage criteria defined in ~\cite{test_coverage_criteria}, whereas RESTCov supports all.
To ensure the validity and comparability of the measurements, we cross-validated the results produced by our tool against those obtained using the original open-source tool on the system evaluated in our earlier work~\cite{bardakci2025testamplificationrestapis}.
This cross-validation confirmed that the computed coverage values were consistent across tools, providing confidence that the measurements reported in this paper are directly comparable to those from our earlier work.

\section{Results}\label{sec:Results}
In this section, we demonstrate our results and give answers to research questions RQ1, RQ2 and RQ3.

\subsection{Coverage Impact (RQ1)}
\textbf{RQ1: }\RQCoverageImpact 

To answer \emph{RQ1}, we analyze the changes in structural API coverage observed after applying LLM-based test amplification to the industrial test suite. 
Coverage is measured using the same metrics adopted in our prior study~\cite{bardakci2025testamplificationrestapis}.

As shown in Table \ref{tab:coverage}, test amplification led to increases across multiple coverage dimensions.
This indicates that the amplified tests exercised API behavior that was not covered by the original test suite.
In particular, amplification improved coverage related to multiple dimensions, including additional operations, parameters, and error conditions.
The results suggest that the LLM is able to explore both alternative execution paths and boundary cases beyond the initial happy-path scenarios.
While the absolute coverage values vary across prompts, the relative improvements indicate that LLM-based test amplification can effectively strengthen API-level test suites in an industrial setting.
Notably, Prompt 2—despite providing additional contextual information via the OpenAPI specification—resulted in lower overall coverage gains than Prompt 1.
A closer inspection suggests that the richer input led the model to focus on localized parameter variations for fewer endpoints, rather than exploring additional operations or execution paths.

Overall, these results show that LLM-based test amplification remains effective at improving structural API coverage under industrial constraints. 
However, the observed coverage gains are influenced by the characteristics of the prompts and the system itself.
The stricter validation rules and state-dependent behavior limit the extent to which additional API elements can be exercised automatically.

\begin{table}[htbp]
\centering
\caption{Structural API coverage before and after test amplification}
\label{tab:coverage}
\begin{tabular}{l c c c c}
\toprule
\multirow{2}{*}{\textbf{Metric}} &
\multirow{2}{*}{\textbf{Baseline}} &
\multicolumn{3}{c}{\textbf{After Amplification}} \\
\cmidrule(lr){3-5}
 &  & \textbf{Prompt 1} & \textbf{Prompt 2} & \textbf{Prompt 3} \\
\midrule
Path          & 25\%   & 50\%   & 25\%   & 100\%  \\
Operation     & 16.7\% & 33.3\% & 16.7\% & 100\%  \\
Status Class  & 8.3\%  & 25\%   & 16.7\% & 100\%  \\
Status Code   & 5.9\%  & 17.6\% & 11.8\% & 70.6\% \\
Response      & 4.2\%  & 8.3\%  & 4.2\%  & 25\%   \\
Request       & 0\%    & 0\%    & 0\%    & 33\%   \\
Parameter     & 9.1\%  & 9.1\%  & 9.1\%  & 45.5\% \\
\bottomrule
\end{tabular}
\end{table}

\subsection{Test Quality and Post-Processing Effort (RQ2)}
\textbf{RQ2: }\RQEffortAndPracticalCost

To answer \emph{RQ2}, we note an example test case and the amount and nature of human intervention required to make the LLM-amplified tests executable and consistent with the existing industrial test suite.
We share the test scenario that is generated in an anonymized form on Listing~\ref{lst:gpt4amplified_test_csharp} and Listing~\ref{lst:gpt4amplified_test_gherkin}.
As described in Section 4, test amplification produced both Gherkin scenarios and corresponding C\# step definitions, and before integration of the code snippets, they were subsequently reviewed and adapted.

\begin{lstlisting}[style=csharpcode, caption={Amplified Test Script C\# Code}, label={lst:gpt4amplified_test_csharp}]
#region POST ITEMS
[When("I create an item with the 
following data:")]
public async Task WhenICreateAnItem(Table table)
{
    var itemData = table.CreateInstance<ItemDto>();
    await Service.CreateItem(itemData);
}

[Then("the response contains the created item with
    Name {string}")]
public async Task 
ThenTheResponseContainsTheCreatedItem(string name)
{
    var item = await GetResponseAs<ItemResponseModel>();
    Assert.Equal(name, item?.Name);
}
#endregion

\end{lstlisting}

\begin{lstlisting}[style=gherkincode, caption={Amplified Test Script Gherkin Code}, label={lst:gpt4amplified_test_gherkin}]
Rule: POST items functionality

Scenario: POST items - HTTP 200
    Given I am successfully authorized
	When I create an item with the following data:
	   | Name      | Number  | Type | Id           |
	   | ItemName  | 123456  | Type | 2c45a, 1b7a8 |
	Then the response code is 200
	And the response contains the created item with Name "ItemName"
\end{lstlisting}

Table \ref{tab:amplification-effort} summarizes the post-processing effort required during this process.
While the LLM generated a substantial number of additional test cases, not all generated artifacts were directly usable.
In particular, there were portions of the generated C\# code that duplicated functionality that already existed in the code base.
These redundant code fragments were discarded, and the amplified Gherkin scenarios were instead connected to existing step definitions.

\begin{table}[htbp]
\centering
\caption{Test amplification output and post-processing effort}
\label{tab:amplification-effort}
\small
\begin{tabular}{l c c c}
\toprule
\textbf{Metric}         & \textbf{Prompt 1} & \textbf{Prompt 2} & \textbf{Prompt 3} \\
\midrule
Generated Tests         & 4 & 5 & 11 \\
Executable Tests        & 4 & 5 & 11 \\
Merged Tests            & 3 & 4 & 11 \\
Discarded Tests         & 1 & 0 & 0  \\
Edited Lines of Code (LoC) & 1 & 2 & 21 \\
\bottomrule
\end{tabular}
\end{table}

Despite this manual intervention, the overall effort remained manageable and localized regardless of the prompting technique used.
Most edits were mechanical and did not require rethinking the test intent or expected behavior.
Importantly, the amplified tests preserved readability and followed the structure of existing BDD-style tests.
Thus, it reduced the cognitive effort required during review.
These findings indicate that, while LLM-based test amplification does not eliminate the need for human involvement, it can be integrated into an industrial testing workflow with a reasonable and predictable level of effort.

\subsection{Observations and Anomalies (RQ3)}
\textbf{RQ3: }\RQObservationsAndLimitations

Applying LLM-based test amplification in an industrial context revealed several practical challenges and behaviors that were not observed or were less pronounced in the previous study~\cite{bardakci2025testamplificationrestapis}. 
These observations highlight differences between controlled experimental settings and real-world testing environments and provide insight into the limitations of directly transferring open-source driven techniques to industry.

A first observation concerns the interaction between generated tests and existing test infrastructure.
LLM was generally able to generate syntactically correct Gherkin scenarios and C\# step definitions.
However, it occasionally assumed the existence of helper methods or abstractions that were not present in the code base while not generating them.
Additionally, sometimes it failed to reuse available shared steps.
This mismatch required manual reconciliation and reflects the difficulty of inferring project-specific testing conventions from limited context.

A second challenge relates to state-dependent and environment-sensitive behavior. 
Several amplified tests relied on implicit assumptions about system state, such as the availability of pre-existing entities or the order of test execution.
In the industrial system, these assumptions sometimes led to flaky or non-deterministic behavior.
Addressing these issues required additional setup or cleanup logic that was not necessary in our previous setting.

A further challenge stemmed from the presence of industrial authentication and authorization mechanisms.
Unlike the open-source system used in our previous study, the industrial API relied on multiple authentication layers, including token management and role-based access control.
Several amplified tests were making minor authentication errors, leading to failures unrelated to the functional behavior under test.
Supporting these mechanisms required additional inspection and increased setup complexity.
Additionally, it reduced the degree of end-to-end automation achievable in practice.

We also observed limitations in the LLMs’ ability to respect domain-specific constraints and business rules. 
In some cases, amplified tests exercised API calls with combinations of inputs that were syntactically valid but semantically unrealistic given the domain logic.
While such tests occasionally revealed anomalies, they also increased the number of generated tests that needed to be filtered or adapted manually.

Finally, the size and complexity of the overall system affected how the OpenAPI specification could be used during test amplification. 
The original OpenAPI specification covered a large number of services and endpoints beyond the scope of the microservice under study.
To make LLM-based test amplification feasible, we manually refined the OpenAPI specification to produce a more compact, focused version that contains only the endpoints and schemas relevant to the selected service.
This additional preprocessing step was necessary to reduce noise in the prompts and to guide the LLM toward generating tests that were applicable in practice.
However, it introduced extra manual effort.
This highlights a challenge for software testing techniques in large-scale industrial systems.

Overall, these observations show that LLM-based test amplification behaves differently in industrial settings than in open-sourced environments. 
While the approach remains useful, its effectiveness depends on careful integration with existing test infrastructure, explicit handling of state and environment assumptions, and realistic expectations regarding automation and human oversight.

\section{Lessons Learned}
Based on our experience applying LLM-based test amplification in an industrial REST API testing context, we present the following lessons.
These lessons reflect practical insights gained during the replication study and are intended to inform both researchers and practitioners considering the adoption of similar techniques in real-world environments.

\subsection{L1: Endpoint complexity matters more than endpoint count}
\paragraph{Observation}
Although the industrial study focused on fewer endpoints than the prior evaluation~\cite{bardakci2025testamplificationrestapis}, the amplified tests exercised substantially more complex behavior, including authentication, state-dependent interactions, and domain-specific validation (e.g., authenticated access with tokens, multi-step workflows relying on pre-existing entities, and input constraints derived from business rules), as discussed in Section~\ref{ref:ServiceUnderTest}.
\paragraph{Explanation}
In industrial systems, individual endpoints often encapsulate richer logic and stricter constraints than those found in open-source testing applications.
As a result, increasing test strength requires exploring deeper behavioral variations of a small number of endpoints rather than broad coverage across many simple ones.
\paragraph{Practical Takeaway}
Practitioners should prioritize selecting representative, behaviorally rich endpoints when applying LLM-based test amplification, rather than attempting to maximize endpoint count.

\subsection{L2: Test amplification must align with existing testing practices}
\paragraph{Observation}
LLM-generated tests were most useful when they adhered to the existing testing style and infrastructure, such as BDD-based Gherkin scenarios and reusable step definitions.
\paragraph{Explanation}
When generated artifacts deviated from established testing conventions, additional manual effort was required to reconcile them with the existing code base.
Conversely, alignment with current practices reduced post-processing effort and improved test acceptability.
\paragraph{Practical Takeaway}
To reduce integration effort, LLM-based test amplification should be guided by explicit prompts and context that reflect the project’s testing framework, language, and architectural conventions.

\subsection{L3: Human review remains essential}
\paragraph{Observation}
While many amplified tests were executable and readable, human intervention was consistently required to discard redundant code, resolve assumptions about system state, and integrate generated tests with existing infrastructure.
\paragraph{Explanation}
Industrial test suites often contain shared abstractions, helper utilities, and implicit assumptions that are difficult for LLMs to infer from limited context.
Human oversight ensures correctness, maintainability, and consistency with organizational standards.
\paragraph{Practical Takeaway}
LLM-based test amplification should be view-\allowbreak\ ed as a developer-assistance technique rather than a fully automated replacement for human-authored tests.

\subsection{L4: Specification-driven amplification requires careful scoping in large systems}
\paragraph{Observation}
The full OpenAPI specification of the industrial system was too large and heterogeneous to be used directly for test amplification.
\paragraph{Explanation}
Providing overly broad specifications introduced noise into prompts and led to irrelevant or inapplicable results.
Manually refining the specification to focus on the service under study was needed.
\paragraph{Practical Takeaway}
In large-scale systems, practitioners should extract and maintain service-specific API specifications when using LLM-based techniques that rely on documentation as input.

\subsection{L5: Research tools require adaptation for industrial use}
\paragraph{Observation}
Existing open-source tools used for evaluating test amplification in previous settings could not be applied directly in the industrial environment.
\paragraph{Explanation}
Differences in documentation styles, infrastructure, and CI pipelines exposed limitations in research tooling that are not apparent in controlled settings.
Developing or adapting tools was necessary to obtain reliable measurements.
\paragraph{Practical Takeaway}
Researchers aiming for industrial adoption should anticipate the need for tool adaptation and validation when transferring experimental techniques to real-world systems.

\section{Threats to Validity}
As with any empirical study conducted in an industrial setting, several factors may threaten the validity of our findings.
We discuss these threats in terms of commonly used validity dimensions and describe the measures taken to mitigate them where possible.

\subsection{Internal Validity}
A potential threat arises from the non-deterministic nature of LLM outputs, which may lead to variability in the generated tests across runs. To mitigate this, we followed a consistent interaction process with the LLM and avoided iterative prompt tuning during result collection.
However, some variability remains inherent to the approach.

Another threat to internal validity stems from the manual post-processing of the generated tests.
One of the authors performed this post-processing after test amplification.
The resulting test code was then submitted as a pull request, reviewed in the same manner as other code changes, and followed the standard development process used within the organization.
Human intervention may introduce bias when deciding which generated tests to retain, modify, or discard.
To reduce this risk, post-processing decisions were guided by explicit criteria, such as executability and conformity with existing test infrastructure, rather than by perceived test quality or expected outcomes.

\subsection{Construct Validity}
\subsection{Construct Validity}
Structural API coverage is useful for assessing how extensively an API is exercised.
However, it does not directly measure fault detection or overall test adequacy.
Similarly, post-processing effort, measured in lines of code, may not fully capture the cognitive effort required to review and integrate generated tests.

To mitigate these threats, we use clearly defined, tool-supported coverage metrics~\cite{RESTCov_2025} and apply the same measurement procedure across all prompts and across both our prior study~\cite{bardakci2025testamplificationrestapis} and the present study.
Moreover, we keep the prompts and evaluation protocol consistent with our prior work to ensure comparability and minimize changes in measurement constructs across settings.

\subsection{External Validity}
External validity concerns the generalizability of the results. Our study focuses on a single industrial partner and a single microservice, which may limit the extent to which the findings generalize to other organizations, domains, or system architectures.
Although the selected endpoints were chosen to be representative of common industrial REST API behaviors, different systems may exhibit different constraints or testing practices.

Furthermore, the study was conducted using a specific LLM setup and tooling configuration, influenced by organizational policies.
Results may differ when using other models, interaction modes, or integration approaches.

\subsection{Conclusion Validity}
A potential threat arises from the use of different tooling for coverage measurement compared to our prior study.
To mitigate this, we cross-validated the coverage results produced by our tool against those obtained with the original open-source tool on the open-source system.
Thus, we increase confidence in the reported measurements.

Another threat concerns the study's limited duration and scope.
Our study may not capture longer-term effects such as test suite evolution or maintenance overhead.
Future longitudinal studies would be needed to assess these aspects more thoroughly.

\section{Conclusion}
In this paper, we report on an industrial replication of LLM-based test amplification for REST APIs.
Building on our earlier open-source example study, we applied the same amplification method to a production microservice at one of the largest logistics company in Belgium.
Despite differences in scale, technology stack, and organizational constraints, our results show that LLM-based test amplification can strengthen industrial API test suites by improving structural coverage and exercising additional API behavior.

From a practitioner’s perspective, our findings suggest that LLM-based test amplification is effective when applied in a focused manner to representative, behaviorally rich endpoints.
While human review and post-processing remain necessary, the required effort was manageable and largely mechanical.
The study also highlights the importance of the amplification technique, particularly in environments that rely on BDD-style testing or strict reuse of test infrastructure.

From a research perspective, this work underscores the gap between open-source example evaluations and industrial realities.
Several challenges observed in this study—such as specification scoping, tooling adaptation, and state-dependent behavior—are unlikely to surface in controlled experimental settings.
Addressing these challenges is essential for translating research on LLM-based test amplification into practical impact.

Looking ahead, future work includes replicating the study across multiple services and organizations to determine whether the approach scales well or can be efficiently adapted across different environments. 
In addition, exploring automated support for specification refinement and ways to reduce human intervention in the process are two important activities to increase the robustness of the approach.
Also, it is worth measuring the fault detection adequacy of LLM-amplified tests, setting up user studies to assess their quality in different dimensions, and studying their adaptation over long periods of time, especially considering long-term maintainability.

\begin{acks}
This work is supported by the Research Foundation Flanders (FWO) via the BaseCamp Zero Project under Grant number S000323N.
\end{acks}


\appendix

\end{document}